\documentclass{aa}

\usepackage{graphics,epsfig}

\newcommand{\kms}{km~s$^{-1}$}
\newcommand{\cm}{cm$^{-2}$}

\newcommand{\dla}{ damped Lyman-$\alpha$ }
\newcommand{\NHI}{N_{\rm HI}}
\newcommand{\noi}{\noindent}

\newcommand{\dV}{\Delta V}

\begin{document}
\title{Detection of a multi-phase ISM at $z=0.2212$.}
\titlerunning{Multi-phase ISM at $z=0.2212$}
\author{Nissim Kanekar \inst{1}\thanks{nissim@ncra.tifr.res.in},
Tapasi Ghosh \inst{2}\thanks{tghosh@naic.edu},
Jayaram N Chengalur\inst{1}\thanks{chengalu@ncra.tifr.res.in}}
\authorrunning{Kanekar, Ghosh \& Chengalur}
\institute{National Centre for Radio Astrophysics, Post Bag 3, Ganeshkhind, Pune 411 007 \and
NAIC, Arecibo Observatory, HC 3  Box 53995, Arecibo, PR00612, USA}
\date{Received mmddyy/ accepted mmddyy}
\offprints{Nissim Kanekar}

\abstract{ We present sensitive Giant Metrewave Radio Telescope (GMRT) and
high-resolution Arecibo HI 21-cm observations of the  damped
Lyman-$\alpha$ absorber (DLA) at $z=0.2212$ towards OI~363 ($B2~0738+313$).
The GMRT and Arecibo spectra are in excellent agreement and yield a spin 
temperature $T_s = 890 \pm 160$~K, consistent with earlier lower sensitivity 
observations of the system. This value of $T_s$ is far higher than spin 
temperatures measured for the Milky Way and local spirals but is similar to $T_s$
values obtained in the majority of damped absorbers ($T_s \ga
1000$~K).\\
The high velocity resolution of the Arecibo spectra enables us to obtain
estimates of physical conditions in the absorbing clouds by fitting 
multiple Gaussians to the absorption profile. The spectra are well fit by a 
three-component model with two narrow and one wide components, with 
temperatures ${T_{k_1}} = 308 \pm 24$~K, ${T_{k_2}} = 180 \pm 30$~K 
and ${T_{k_3}} = 7600 \pm 1250$~K, respectively. The last of these is in excellent 
agreement with the expected temperatures for the WNM ($5000 - 8000$~K). Further, the 
mere fact that components are seen with lower temperatures than the 
estimated $T_s$ implies that the absorber must have a multi-phase medium. \\
We use the measured 21-cm optical depth and the above estimates of the kinetic 
temperature to obtain the HI column density in the various components. 
The total column density in the narrow components is found to be 
$\NHI ({\rm CNM}) \le 1.9 \pm 0.25 \times 10^{20}$~\cm,  while that in the 
wide component is $\NHI ({\rm WNM}) \ge 1.26 \pm 0.49 \times 10^{21}$~\cm. Thus, 
the WNM contains at least $75$~\% of the total HI in the $z = 0.2212$ DLA, unlike 
our Galaxy, in which the CNM and WNM have equitable contributions. 
As conjectured earlier (Chengalur \& Kanekar 2000), this accounts for the 
difference in the 
spin temperatures of the $z = 0.2212$ system and local spirals, suggesting that the 
DLA is probably a dwarf or LSB type galaxy; this is also in agreement 
with optical studies (Turnshek et al. 2001). Finally, the total column density in 
the DLA is found to be $\NHI \sim 1.45 \pm 0.49 \times 10^{21}$~\cm, which agrees 
within the errors with the value of $\NHI = 7.9 \pm 1.4 \times 10^{20}$~\cm, 
obtained from the Lyman-$\alpha$ profile (Rao \& Turnshek 1998). This reinforces 
our identification of the wide and narrow components as the WNM and CNM respectively. 
\keywords{galaxies: evolution: --
          galaxies: formation: --
          galaxies: ISM --
          cosmology: observations --
          radio lines: galaxies}
}
\maketitle
\section{Introduction}
\label{intro}

	Extragalactic gas clouds which lie along the line of sight
to a distant quasar give rise to absorption lines in the quasar spectrum.
For clouds with neutral hydrogen column density $\NHI \ga 
10^{20}$~cm$^{-2}$, the width of the absorption line corresponding to the 
Lyman-$\alpha$ transition is very large because the optical depth is 
substantial even in the Lorenzian wings of the line profile. Essentially, 
at these column densities, the shape of the absorption profile is 
determined not by the Doppler motions of the atoms in the cloud, but 
by the natural profile of the line. These high column density systems 
(called damped Lyman-$\alpha$ absorbers, or DLAs) are rare -- the 
probability of finding a 
DLA in a random search towards a high redshift quasar is $\sim 0.25$ per unit 
redshift interval (\cite{storrie}). Nonetheless, DLAs contain the bulk of 
the observed neutral gas at high redshift ($z \sim 3$) and are hence logical 
candidates for the precursors of today's galaxies.

	Since quasars are essentially point-like objects in the optical 
and UV wavebands, their absorption spectra in these bands contain 
information only on the gas illuminated by the narrow pencil beam 
of continuum emission from the quasar. It is largely for this reason that 
the typical transverse size, luminosity and mass of DLAs remain controversial
despite almost two decades of systematic study. Damped systems at high redshift 
have traditionally been assumed to be disk galaxies (e.g. \cite{wolfe88}). 
Some support for this hypothesis is provided by high spectral 
resolution optical studies of the absorption profiles of low ionization metals 
in the absorbers. These metals have ionization potentials lower than that of 
HI, as well as unsaturated line profiles; these profiles thus trace
the kinematics of the neutral gas in the DLA (\cite{lanzetta92}). Prochaska 
\& Wolfe (1997, 1998; but see Ledoux et al. 1998) used a large sample 
of DLAs to argue that the metal line profiles were consistent with those 
arising from rapidly rotating massive disks. However, it was later shown that 
such profiles could also be generated in a variety of DLA models, ranging from 
merging sub-galactic blobs (\cite{haehnelt98}), to randomly moving clouds 
in a spherical halo (\cite{mcdonald99}). At low redshifts, Hubble Space Telescope 
(HST) and ground-based imaging and spectroscopy can be used to carry out detailed 
studies of the galaxies resposible for the damped absorption. These 
observations indicate that DLAs are associated with a wide variety of galaxy 
types and are not exclusively (or even predominantly) spiral galaxies 
(\cite{lebrun97,rao98,raovla}). In this paper, we will discuss HI 21-cm observations 
of the DLA at $z=0.2212$ toward the radio-loud quasar OI~363 ($B2~0738+313$); this
absorber has been identified by recent studies as a dwarf, 
or possibly LSB-type, galaxy, with luminosity $L \sim 0.1 L_{*}$ 
(\cite{turnshek01,cohen01}). 

HI 21-cm absorption was first detected in the DLA at $z=0.2212$ by Lane et al. (1998),
using the Westerbork Synthesis Radio Telescope (WSRT); these observations,
however, only marginally resolved the line. A higher sensitivity, higher resolution 
spectrum using the Giant Metrewave Radio Telescope (GMRT) was presented by Chengalur 
\& Kanekar (1999). This 21-cm profile showed a single, narrow component with
FWHM $\sim 5.5$~\kms~and was  used to estimate the spin temperature 
of the absorbing gas. This was found to be quite high, $T_s = 1120 \pm 200$~K. 
In the Galaxy, gas with a temperature of $\sim 1000$~K is unstable. In standard
models of the galactic interstellar medium (ISM) (\cite{wolfire95,kulkarni88}), 
neutral hydrogen exists in only two stable phases, a cold phase at $\sim 80$~K 
and a warm phase at $\sim 8000$~K. For a multi-phase medium, however, the measured 
spin temperature is the column 
density weighted harmonic mean of the temperatures of the different phases. 
It was hence proposed that the high values of $T_s$ estimated for this and other DLAs 
were due to a substantial fraction of their neutral gas being in the warm phase 
(\cite{chengalur00,kanekar01}).

	  Here, we present high-resolution Arecibo HI 21-cm observations 
of the $z=0.2212$ absorber, supplemented by a deep GMRT 21-cm observation 
of the system. These observations enable us to directly determine the 
kinetic temperature of the absorbing gas and thus, to distinguish between 
the warm and cold phases. We find, as conjectured previously, that the bulk 
of the gas is indeed in the warm phase, which contains at least three-fourths of 
the total HI along this line of sight. This is the second DLA for which it has
been observationally established that a high spin temperature is due 
to a preponderance of gas in the warm neutral medium; Lane et al. (2000) 
show that this phase contains at least two-thirds of the gas in a DLA at $z=0.0912$ 
(also, coincidentally, towards OI~363).  

	The rest of the paper is organised as follows: the GMRT and Arecibo 
observations and data analysis are described in Sect.~\ref{sec:obs}. 
Sect.~\ref{sec:anal} presents the absorption spectra; the spin temperature 
of the absorber and the temperatures of the different absorbing components 
are also estimated here. Finally, Sect.~\ref{sec:dis} discusses our results 
both in the context of observations of the $z = 0.2212$ DLA at other wavelengths, 
and with respect to their general implications for the nature of systems which give rise 
to damped Lyman-$\alpha$ absorption. 

\section{Observations and data analysis}
\label{sec:obs}

\subsection{Arecibo observations and data analysis.}
\label{ssec:arecibo}

	The $z = 0.2212$ absorber towards OI~363 was observed on a number 
of occasions in February, April and August 2000, using the 305-m Arecibo 
radio  telescope. All observations  were done in total power mode, with 
the ``L-wide'' receiver, using 9--level sampling for the auto-correlation 
spectrometer. Two orthogonal circular polarization channels were observed 
simultaneously. The first two sessions (in February and April
2000) used bandwidths of 3.125 and 6.25 MHz, centred at a heliocentric 
redshift of $z = 0.2212$, and divided into 2048 channels. This yielded 
velocity resolutions of $0.3935$ and $0.79$~\kms~respectively. 

	OI~363 has substantial continuum flux density ($\sim 2$~Jy) at L~band and the
bandpass is hence dominated by systematics due to standing wave patterns.
To improve the bandpass calibration, the observations were carried out in 
a ``double switched'' mode. OI~363 was observed first, followed by an 
observation of blank sky. Next, a similar position-switched observation
was made for a nearby source, OI~371 (B~0742+318, flux density $\sim$~1.4~Jy).
OI~363 and OI~371 have very similar  
declinations and continuum flux densities; the observations were so timed that 
the alt-az track of the feed was almost the same  for all phases of the observing
cycle. Each of these phases was $\sim 4$~minutes long (with 5-second data records) and this 
cycle was repeated on each observing day for as long as the sources were visible from Arecibo. 
All observations were carried out at night, to minimise solar effects.
The total on-source time for OI~363 was about 9 hours.

	A second set of observations was carried out in August 2000, with 
the aim of better resolving the narrow absorption components seen in the 
initial Arecibo observations. Far smaller bandwidths of 0.781 and 0.195~MHz were 
hence used, again divided into 2048 channels, and with two polarizations 
at each setting. This gave velocity resolutions of 
$\sim 0.099$ and $\sim 0.025$~\kms~respectively. These observations 
were carried  out in the standard on-off position switching mode (i.e.
a single blank sky spectrum was used to calibrate the bandpass), since 
the wider of the two bandwidths (0.781 MHz) was narrower in frequency
than one cycle of the standing 
wave ripple (which is $\sim 1$~MHz at Arecibo). Each on-off cycle was of 
five minutes duration, sub-divided  into 1-second records. The total 
on-source time was $\sim$ 35 minutes.

	The data for both sets of observations were reduced using the 
Arecibo software package {\sc Analyz}. For the February and April 
observations, each four-minute spectrum was initially inspected for radio 
frequency interference (RFI). If interference was seen, the individual 
5-sec records of the four-minute run were passed through a standard 
RFI excision program, and all records with strong ($ > 10 \sigma$) features
were removed.  Each four-minute spectrum was also inspected for the 
presence of standing waves. It was  found that the fluctuations on each 
such position-switched spectrum were indeed dominated by standing waves 
across the bandpass. However, when an OI~363 spectrum was divided 
by the corresponding spectrum of OI~371 using the formula, 

\begin{equation} 
{\rm Spectrum} = \left[ \frac{ {\rm On - Off}}{{\rm Off}} \right]_{{\rm OI~363}} {\Bigg/}
 \left[ \frac{ {\rm On - Off}}{{\rm Off} } \right]_{{\rm OI~371}} \; 
\end{equation}

\noi the resultant spectrum contained essentially Gaussian noise. This ratio of the two 
spectra was also bandpass corrected and had no remaining effects  
from the azimuth/zenith-angle dependence of gain. It was then multiplied by the
flux density of OI~371 (taken to be $1.4$~Jy at 1160~MHz) to convert the 
spectrum into Jy.

	The basic data editing for the August spectrum was carried out 
in a similar manner.  Here, however, the individual five-minute spectra 
were obtained by the formula 

\begin{equation}
{\rm Spectrum} = { \left[ \frac{ {\rm On - Off}}{{\rm Off}} \right]_{{\rm OI~363}} } \; \; 
\end{equation}

\noi The spectra were then corrected for the System Equivalent Flux Density (SEFD) v/s Zenith angle 
dependence, to obtain the final spectra in Jy. 

For all data sets, the above procedure was carried out separately for the two polarizations. 
Individual four- and five-minute spectra were averaged together (using the appropriate 
weights) to produce the final spectrum for each bandwidth. A linear baseline 
was then fitted to these spectra (excluding the location of the line) and subtracted.
Finally,  the spectra were divided by the mean continuum flux density to convert
them into optical depth (since the line is optically thin, a simple
division by the continuum level suffices to convert from flux density into
optical depth).

\subsection{GMRT observations and data analysis.}
\label{sec:GMRT}

GMRT observations of the $z = 0.2212$ absorber were carried out on
19 October, 2000. The standard 30--station FX correlator, which gives 
a fixed number of 128 channels over a bandwidth which can be varied between 
64 kHz and 16 MHz, was used as the backend. A bandwidth of 1 MHz was used 
for the observations, yielding a spectral resolution of $\sim 2.0$~\kms.
3C147 and 3C286 were used to calibrate the absolute flux density scale and system bandpass; 
no phase calibrator was used since OI~363 is itself unresolved by the GMRT. 
The total on-source time was around four and a half hours, with sixteen 
antennas present. 

The data were analysed in AIPS using standard procedures. The analysis 
was fairly straightforward since there is negligible extended emission 
in the OI~363 field. Continuum emission was subtracted by fitting a 
linear baseline to the U-V visibilities, using the AIPS task UVLIN. 
The continuum subtracted data were then mapped in all frequency channels and a spectrum 
extracted at the quasar location from the resulting three-dimensional 
data cube. Spectra were also extracted at other locations in the cube to 
ensure that the data were not corrupted by interference. A spectrum was 
also obtained by simply vector averaging the U-V visibilities together, 
using the AIPS task POSSM. Since the two methods are essentially equivalent 
for the present case of a point source at the telescope phase centre, we  
use the POSSM spectrum in the discussion below. The RMS noise on this spectrum 
is $\sim 2.8$ mJy, while the flux density of OI~363 was measured to be 2.25 Jy. 
Our experience with the GMRT indicates that the flux density calibration is reliable to
$\sim 15$\%, in this observing mode.

\section{The temperature of the absorbing gas.}
\label{sec:anal}
	
\begin{figure*}[t!]
\begin{center}
\epsfig{file=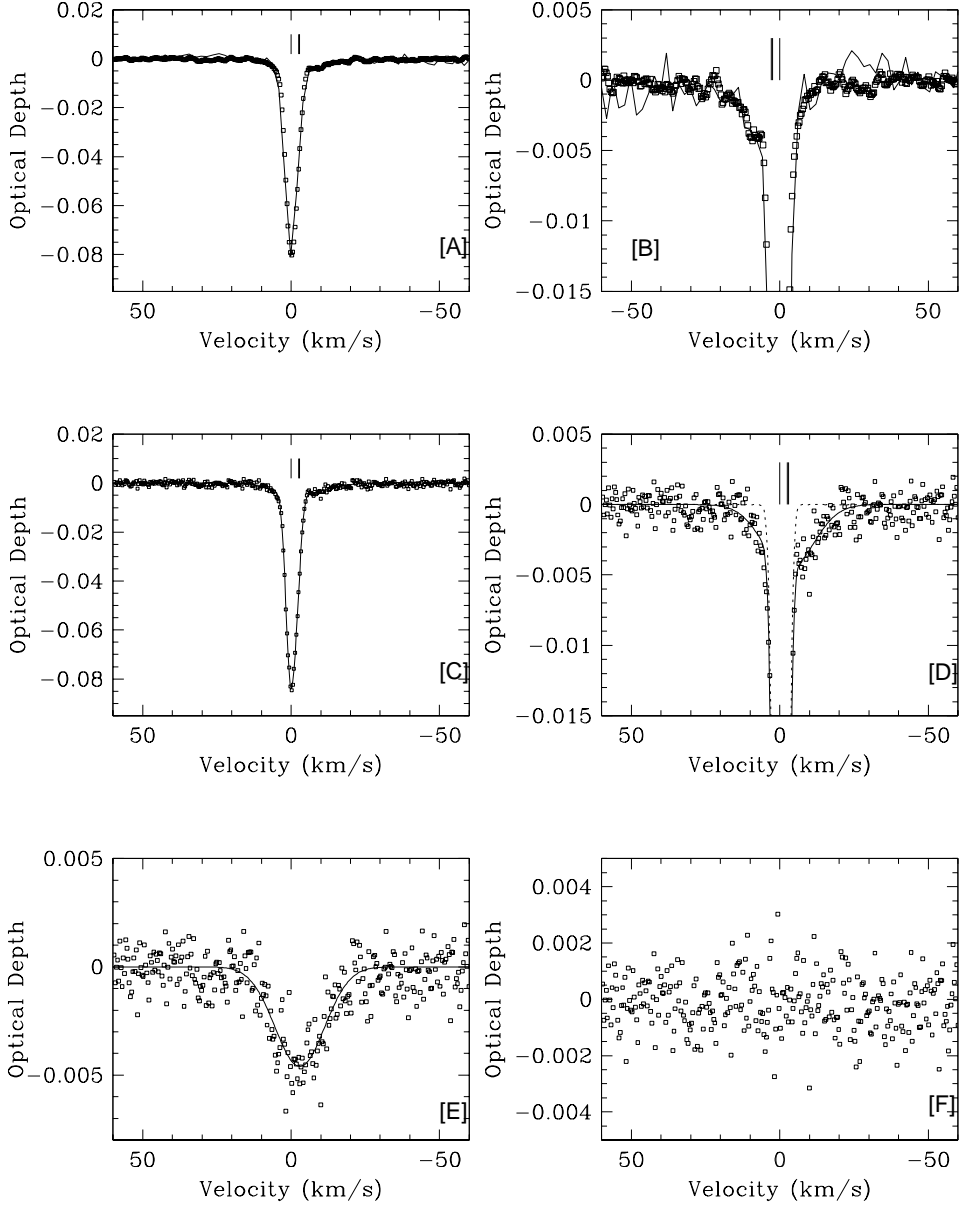,height=7truein}
\end{center}
\vskip 0.0 in
\caption{ The 21-cm absorption spectra of the $z=0.2212$ absorber towards OI~363.
{\bf [A]}~A comparison of the Arecibo 0.4-\kms~resolution spectrum (open
squares) with the GMRT 2.0-\kms~resolution spectrum (shown as a thin solid 
line). The Arecibo spectrum has been smoothed to a resolution of $\sim 2$-\kms~for 
the comparison. The vertical lines above $\tau = 0$, indicate the locations of the 
centres of the three components.
{\bf [B]}~The 0.4-\kms~resolution Arecibo spectrum (open squares) shown
superposed with the GMRT 2.0-\kms~resolution spectrum (solid line). This is a zoomed-in
version of panel~[A] to bring out the shallow absorption wing seen in both
spectra. The Arecibo spectrum has again been smoothed to a resolution of $\sim 2$~\kms~for 
the comparison.
{\bf [C]}~The 0.4-\kms~resolution Arecibo spectrum (open squares) with the
three-component Gaussian fit shown as a thin solid line. (see the text and
Table~\ref{tab:Gauss} for the fit parameters).
{\bf [D]}~The 0.4-\kms~resolution Arecibo spectrum (open squares) shown
superposed with the multi-Gaussian fit (thin solid line). This is a zoomed-in 
version of panel~[C] to bring out the shallow absorption wing. The dotted line 
is a two-component Gaussian fit, which clearly brings out the necessity of 
including a wide component.
{\bf [E]}~Residuals left after subtracting the two narrow components from the 
0.4-\kms~resolution Arecibo spectrum (open squares). The fit to the wide 
component (see text and Table~\ref{tab:Gauss}) is shown as a thin solid line.
{\bf [F]}~Residuals left after subtracting the two narrow components and the
wide component from the 0.4-\kms~resolution Arecibo spectrum. }

\label{fig:spectra}
\end{figure*}

	The spectra are shown in Fig.~\ref{fig:spectra} and Fig.~\ref{fig:high}. 
Fig.~\ref{fig:spectra}[A] presents the final Arecibo 3.125-MHz ($\sim 0.395$~\kms~resolution) 
spectrum in open squares. This has been smoothed to a resolution of
$\sim 2$~\kms~to compare it to the GMRT spectrum, shown here as a thin solid line. 
The agreement between the two is excellent. The central narrow absorption component 
can be seen to be slightly asymmetric in both spectra. Further, in addition to the 
deep narrow component, the 21-cm spectrum also has a shallow broad component, seen 
in both the Arecibo and the GMRT spectra.  This component can be more clearly seen in the 
zoomed-in plot of Fig.~\ref{fig:spectra}[B], where the two  spectra 
have again been superposed; again, the solid line is the GMRT spectrum, 
while the open squares are the points on the Arecibo 3.125-MHz spectrum.
Fig.~\ref{fig:spectra}[C] shows the Arecibo 3.125-MHz spectrum at the original 
resolution of $\sim 0.4$~\kms~(open squares); the thin solid line is the multi-Gaussian 
fit to the spectrum which will be discussed later. This spectrum has an RMS
noise of $\sim 0.001$ in optical depth per $\sim 0.4$~\kms~channel (i.e. $\sim 2$~mJy). 
The deepest absorption occurs at a heliocentric frequency of 1163.075 MHz, i.e. at 
a heliocentric redshift of $z = 0.221250 \pm 0.000001$.

\subsection{The spin temperature}
\label{sec:t_spin}

        For optically thin absorption by a homogenous HI cloud, the 21-cm 
optical depth $\tau_{21}$ is related to the column density $\NHI$ and spin temperature $T_s$ by

\begin{equation}
\NHI = { 1.8\times 10^{18} T_s \over f} \int \tau_{21} dV
\label{eqn:nh-ts}
\end{equation}

\noindent where $f$ is the covering factor of the absorbing gas.  OI~363 is
a highly compact radio source (\cite{lane00,stanghellini97}) and, as discussed in Chengalur 
\& Kanekar (1999) and Lane et al. (2000), the covering factor of the absorber 
is likely to be close to unity. As mentioned in Sect.~\ref{intro},  the 
earlier, lower sensitivity GMRT observations of Chengalur \& Kanekar (1999) 
yielded a high spin temperature $T_s = 1120 \pm 200$~K for the absorber. The 
deeper Arecibo observations can be used to obtain 
a more precise estimate of the spin temperature of the system. The 
3.125-MHz Arecibo spectrum was used for this purpose as this, of all the Arecibo spectra, 
provides the optimum balance between resolution and signal-to-noise ratio. 
The integrated optical depth of the spectrum in Fig.~\ref{fig:spectra}[C] 
gives an HI column density of $\NHI = 0.889 \pm 0.007 \times 10^{18}~T_s$~\cm~
for the absorbing gas. The column density obtained from the HST Lyman-$\alpha$
profile is $\NHI = 7.9 \pm 1.4 \times 10^{20}$ \cm~ (\cite{rao98}). Comparing
the two yields a spin temperature $T_s = 890 \pm 160$~K, which agrees
within the errors with the values obtained earlier by Lane et al. (1998)
and Chengalur \& Kanekar (1999).

        In the Galaxy, the neutral ISM has no stable phase with a temperature 
of $\sim 1000$~K (although recent Arecibo observations indicate that a 
significant fraction of neutral gas may be in the unstable phase with 
$500$~K~$<T_k < 5000$~K; \cite{heiles00,heiles01}). Instead, there exists 
a cold phase (the Cold Neutral Medium -- CNM) with a temperature of $\sim 80$~K and 
a warm phase (the Warm Neutral Medium -- WNM) with a temperature of $\sim 8000$~K. For each
individual phase, the spin temperature is essentially the same as the kinetic
temperature of the gas. However, the {\it measured} spin temperature 
for a multi-phase absorber is the column density weighted harmonic mean 
of the temperatures of the different phases (provided that the optical depth 
of the gas in each phase is small; see e.g. Kulkarni \& Heiles 1988). 
The high spin temperature of the $z=0.2212$ DLA suggests that the bulk of 
its neutral gas is in the warm phase, if it has a multi-phase ISM similar 
to that of the Galaxy.

\subsection{Temperatures of the individual phases: detection of the WNM}
\label{sec:wnm}

The high resolution and sensitivity of the Arecibo observations allow us to 
resolve the HI absorption profile into its individual components. As the measured optical depths are much 
smaller than unity, this can be 
used to obtain information on the temperatures of the different phases 
of neutral hydrogen in the absorber. Optically thin absorption lines have Gaussian shapes if the 
absorbing gas has a thermal distribution. 
The kinetic temperature $T_k$ of the 
gas is then related to the velocity width of the absorption profile by 

\begin{equation}
\label{eqn:tk-dv}
T_k = 21.87 \times {\Delta} V^2 \;\; {\rm K}\; ,
\end{equation}

\noindent where $\Delta V$ is the FWHM of the Gaussian line in~\kms. In fact,
the measured velocity width provides an upper limit to the kinetic temperature, 
even in cases of non-thermal contributions to the line width, since the measured 
width must be {\it larger} than (or equal to) the width due to thermal 
motions. Thus, given sufficient velocity resolution, one can obtain the kinetic 
temperatures (or upper limits on $T_k$) of the different absorption components, and  
then, using equation~(\ref{eqn:nh-ts}), their HI column densities (since 
$T_s \approx T_k$, see Kulkarni \& Heiles 1988). This gives an independent 
estimate of the HI column density of the absorber, which can be compared 
to that obtained from the Lyman-$\alpha$ profile.

	Since the narrow component seen in the Arecibo and GMRT spectra 
is asymmetric and there is additional broad weak absorption, we attempted 
to simultaneously fit three Gaussians to the absorption profile. The 3.125-MHz 
Arecibo spectrum was again used for this purpose. (Attempts were also made to 
fit only two Gaussians to the spectrum but, as expected, this was found to leave 
large residuals.) The three component fit to the 3.125-MHz spectrum decomposes 
the central feature into two components, the first with a velocity width 
(FWHM) $\dV_1 = 3.76 \pm 0.12$~\kms~and peak optical depth $\tau_1 = 0.076 
\pm 0.002$, and the second with $\dV_2 = 2.87 \pm 0.25$~\kms~and 
$\tau_2 = 0.022 \pm 0.004$. The  third, wide component has an FWHM of
$\dV_3 = 18.65 \pm 1.11$~\kms~and a peak optical depth $\tau_3 = 0.0046 \pm 0.0004$.
This best-fit model is superposed on the Arecibo 3.125-MHz spectrum in 
Figs.~\ref{fig:spectra}[C] and [D]; here, the fit is shown as a solid line, while 
the open squares indicate the original spectrum. Fig.~\ref{fig:spectra}[E] 
shows the spectrum after subtracting the two Gaussians fit to the narrow 
central component; the third, wide component is clearly visible. This spectrum 
also shows the fit to the wide component (solid line), which appears quite reasonable. 
Finally, Fig.~\ref{fig:spectra}[F] shows the spectrum after subtracting all 
three Gaussian components. The residuals are seen to lie within the noise.

\begin{figure}
\begin{center}
\epsfig{file=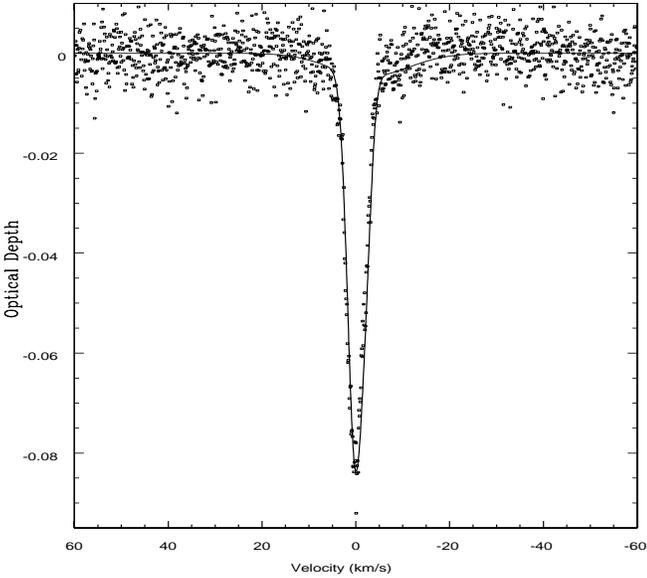,height=3.25truein,width=3.5truein}
\end{center}
\caption{ High resolution ($\sim 0.1$~\kms) Arecibo HI 21-cm spectrum of 
the $z=0.2212$ absorber towards OI~363 (open squares). The thin solid line 
is the fit derived from the $\sim 0.4$-\kms~Arecibo data.}
\label{fig:high}
\end{figure}

\begin{table*}
\begin{centering}
\vskip 0.2 in
\begin{tabular}{|ll|c|c|c|}
\hline
&&&&\cr
\multicolumn{2}{l}{Component} & 1 & 2 & 3 \cr
&&&&\cr
\hline
&&&&\cr
Optical depth &AO Low & $0.076 \pm 0.002$ & $0.022 \pm 0.004$ & $0.0046 \pm
0.0005$ \cr
              & GMRT  &                   &                   & $0.0055 \pm
0.0005$ \cr
              &AO High& $0.080 \pm 0.002$ & $0.016 \pm 0.007$ &
\cr
&&&&\cr
FWHM (\kms)   &AO Low & $3.76 \pm 0.12$   & $2.87 \pm 0.25$   & $18.65 \pm
1.11$    \cr
              &GMRT   &                   &                   & $18.77 \pm
1.82$    \cr
              &AO High& $3.85 \pm 0.30$   & $1.75 \pm 0.69$   &
\cr
&&&&\cr
$T_k$ (K)     &       & $308 \pm 24$      & $180 \pm 30$      & $7600 \pm 1250$
\cr
&&&&\cr
$\NHI$ ($10^{20}$\cm)&& $1.7 \pm 0.24$    & $0.21 \pm 0.1$    & $12.6 \pm 4.9$
\cr
&&&&\cr
Redshift ($\pm 10^{-6}$) && 0.221250   & 0.221240   & 0.221239 \cr
&&&&\cr
\hline
\end{tabular}
\vskip 0.2in
\caption{\label{tab:Gauss} Parameters of the multi-Gaussian fits to the
different spectra.
AO~Low is the Arecibo 3.125-MHz bandwidth spectrum, GMRT is the GMRT 1.0-MHz
bandwidth spectrum, and AO~High is the Arecibo 0.78125-MHz bandwidth 
spectrum. The kinetic temperature, $\NHI$ and redshift are computed from the 
AO~Low spectrum, which has by far the best signal-to-noise ratio.}
\end{centering}
\end{table*}

The 0.781-MHz bandwidth Arecibo spectrum (shown in Fig.~\ref{fig:high}) has 
far lower sensitivity than the 3.125-MHz spectrum (owing to a much shorter 
integration time). It is consequently not very sensitive to the wide component. 
An independent check of the fit to the narrow components can however be made 
by fitting solely to these components, while keeping the parameters of the 
wide component fixed at the values determined from the 3.125-MHz spectrum. 
The parameters obtained from this fit agree (within the errors) with those 
listed above for the 3.125-MHz spectrum although there is a suggestion that the
secondary narrow component may be even narrower than determined from the
latter. Similarly, because of its relatively poor resolution, the GMRT 
spectrum is not particularly sensitive to the decomposition of the narrow
component. However, the fit to the wide component (as obtained after keeping the
parameters of the narrow components fixed to the values obtained from the 
3.125-MHz Arecibo spectrum) is in excellent agreement with that obtained
independently from the Arecibo data. Table~\ref{tab:Gauss} summarises the parameters 
of the fits to the different spectra.

Equation~(\ref{eqn:tk-dv}) can now be used to obtain the kinetic temperatures
of the individual Gaussian components, using the velocity widths listed in
Table~\ref{tab:Gauss}. This yields ${T_{k_1}} = 308 \pm 24$~K and 
${T_{k_2}} = 180 \pm 30$~K, for the two narrow components, and 
${T_{k_3}} = 7600 \pm 1250$~K, for the broad component. Clearly, the measured 
spin temperature $T_s = 890 \pm 160$~K is quite different from the 
physical temperatures of the absorbing clouds. The kinetic temperatures of
the first two components are larger than those of cold clouds in the 
Milky Way (\cite{braun92}), while the temperature of the third ($T_k\sim $ 7600~K) 
is in excellent agreement  with estimates of the temperature of the WNM 
($\sim 5000 - 8000$ K; Kulkarni \& Heiles 1988, Carilli et al. 1998). Note
that recent Arecibo observations of the WNM in the Milky Way have found 
some evidence  for {\it lower} temperatures than the above, with $T_k \sim 2000$~K
(\cite{heiles00,heiles01}), although these results are still in preliminary
form. 

	These estimates of the kinetic temperature can be used in 
equation~(\ref{eqn:nh-ts}) to obtain the HI column densities of the 
different components. This yields $\NHI (1) = 1.7 \pm 0.24 \times 10^{20}$~\cm~ 
and $\NHI (2) = 2.1 \pm 1.0 \times 10^{19}$~\cm~ for the two narrow components, 
and $\NHI (3) = 1.26 \pm 0.49 \times 10^{21}$ \cm~ for the wide component. 
The total column density in all three components is $\NHI = 1.45 \pm 0.49$~\cm, 
larger than (but, given the errors, still in agreement with) the value of 
$7.9 \pm 1.4 \times 10^{20}$~\cm~obtained from the Lyman-$\alpha$ line (\cite{rao98}). 
Given that the 21-cm $\NHI$ measurements are, strictly speaking, upper 
limits, since there may be non-thermal contributions to the line width, the agreement 
between the column densities obtained from the 21-cm profile and the Lyman-$\alpha$ 
profile is quite good. Thus, the 21-cm profile of this absorber agrees fairly
well with what would be expected from an absorber with a multi-phase medium 
much like that of the Galaxy (at least, in terms of the {\it temperatures} of the two 
phases).


	It should be noted that non-thermal motions of a given magnitude will make 
a larger difference to narrow lines than to wide lines. It is therefore likely that 
the column density in the narrow components (which we identify with the CNM) 
are more severely over-estimated.  The total contribution of the CNM to the 
HI column density is  $\NHI ({\rm CNM}) \le 1.9 \pm 0.25 \times 10^{20}$ \cm, i.e. 
at most 24\% of the total column density obtained from the \dla line (\cite{rao98}). 
Of course, it is possible that the wide component also has a low kinetic temperature 
and that its large width is primarily due to non-thermal motions of the absorbing gas. 
For example, this could be due to the blending of a number of narrow CNM lines of low optical 
depth, rather than due to absorption by the WNM. Even in such a situation, however, 
the total contribution of such clouds to the HI column density would be very low, 
leaving unaccounted a substantial fraction of the column density, which must lie 
in a still hotter component. For example, if we assume that the hypothetical blended 
cold clouds have 
kinetic temperatures $T_k \sim 100$~K, then the total HI column density in the wide 
component is negligible, $\NHI \sim 1.7\times 10^{19}$~\cm. The bulk of the gas seen in
the Lyman-$\alpha$ profile would hence still be required to be in a phase whose 
temperature is sufficiently large to make the 21-cm optical depth lower than our 
detection threshold. Thus, regardless of whether or not the wide component is 
identified with the WNM, one requires $\sim 76\%$ (or possibly, even more) of the HI 
to be contained in the WNM. Indeed, the mere fact that we observe components 
in the absorption profile whose kinetic temperatures are smaller than the average
spin temperature of the gas means that the gas must have multiple phases, including
a warm phase. Given the relatively good agreement between the \dla and 21-cm 
HI column densities and the reasonable value of the derived kinetic temperature 
of the wide component, we consider it highly likely that the wide component is indeed
gas in the WNM phase. Note that in calculating the fraction of gas in the WNM phase
we have  assumed that the \dla $\NHI$ measurement is the best estimate of the
total neutral-hydrogen column density.

\section{Discussion}
\label{sec:dis}

	As discussed in the previous section, the 21-cm absorption profile of the
$z=0.2212$ DLA is in excellent agreement with that expected were the 
absorption to arise in a multi-phase medium similar to that of the Milky Way. 
However, unlike the Galaxy, where the CNM and WNM both make equitable 
contributions to the total HI column density (\cite{kulkarni88}), $\sim 75\%$ 
of the neutral hydrogen along this line of sight through the DLA must be in the 
warm phase, i.e. with temperature $\sim 8000$~K, in order to account for its high estimated 
spin temperature.

	The average spin temperature for the $z = 0.2212$ absorber of $\sim 900$~K is
far higher than the typical spin temperatures of $100 - 200$~K found in the Galaxy
and nearby spirals (\cite{braun92, braun97}). High spin temperatures of similar 
magnitude were earlier obtained by Carilli et al. (1996) in DLAs at high redshift. It 
was suggested there that the high $T_s$ values at high $z$ might be explained by 
evolutionary effects. Since then, however, there has been a substantial increase 
in the number of DLAs with 21-cm  observations 
(\cite{lane98}, \cite{chengalur99, chengalur00,kanekar01}) and, as Chengalur 
\& Kanekar (2000) point out, high spin temperatures appear to be typical for 
DLAs at all redshifts. So far, the only DLAs which show low $T_s$ 
values are those known to be associated with the disks of spiral galaxies. Chengalur 
\& Kanekar (2000) (see also Kanekar \& Chengalur 2001) suggested that the high-$T_s$ 
DLAs were likely to be associated with dwarf or LSB-type galaxies, where a 
combination of low central pressures, low dust content and low metallicities result 
in a smaller fraction for the CNM.  Our conclusion that $\sim 75\%$ of the gas along 
the line of sight through the $z = 0.2212$ DLA is in the WNM phase is in good 
agreement with the observations of Young et al. (2000), who find that nearby dwarf 
galaxies have $\sim 80\%$ of their neutral gas in the WNM phase. 

	Apart from possible selection effects, the association of DLAs with
dwarf galaxies is somewhat surprising, both because of the existing paradigm of
DLAs being associated with massive rotating disks
(\cite{prochaska97,prochaska98}),  and the expectation (based on a census of
the HI content of $z=0$ optically catalogued galaxies by \cite{rao93}) that the
bulk of the neutral HI at low redshifts is in large spiral galaxies. However,
blind searches for HI emission at low redshift (i.e.  unbiased by the presence
of a catalogued galaxy; \cite{schneider98, rosenberg00}) also
indicate that there could be substantial amounts of HI in optically faint
galaxies, although these results are still controversial (e.g. see Zwaan et al.
(1997), for an opposing point of view).

	In this context, it is of interest that the three lowest redshift 
DLAs known have all been (tentatively) identified as dwarf (or LSB) galaxies 
(\cite{bowen01,turnshek01,cohen01}) (where, following 
Turnshek et al. 2001, we only consider systems to be DLAs if they meet the 
``classical'' selection criterion for a DLA, i.e. for which an observed 
Lyman-$\alpha$ profile yields an HI column density $\NHI \ge 2\times 10^{20}$~\cm). 
In the case of the current absorber, Le Brun et al. (1997) suggested that a galaxy 
at an impact parameter of $\sim 6^{''}$  from OI~363 was likely to be the DLA host. 
The spectrum of this galaxy (\cite{cohen01}) confirms that it is indeed at the 
correct redshift to produce the damped absorption. Turnshek et al. (2001) present 
both detailed multi-wavelength photometry and optical spectroscopy of this system, 
and find that its colours are consistent with a dwarf ($L \sim 0.1L_*$) early-type 
galaxy. The radial profile, however, indicates the presence of both a bulge and a disk. 
Hence, these authors suggest that this galaxy could be the equivalent of the dwarf spirals 
seen at low redshifts (\cite{schombert95}) and/or might have evolved from the faint blue
galaxies seen at higher redshifts. It should also be noted that the impact parameter 
($\sim 18$)~kpc is somewhat large for a dwarf galaxy and it is thus also possible that 
the absorption arises in an even fainter companion galaxy.

	The present detection of the WNM in the $z=0.2212$ absorber towards OI~363 
is the second case of evidence for a multi-phase medium in an extragalactic 
system. Lane et al. (2000) found that the $z=0.0912$ absorber towards the 
same quasar also has a multi-phase medium, with at most $\sim$ one-third of 
the gas in the CNM phase. Interestingly, this DLA is also likely to be associated with 
a dwarf galaxy; Turnshek et al. (2001) place an upper limit of $\sim 0.1~L_*$ on 
its K band luminosity. It is possible that such systems dominate current samples 
of DLAs because the obscuration for lines of sight passing through gas with both 
high metallicity and high dust content might well be sufficient to make 
medium resolution spectroscopy of the background quasar extremely difficult 
(\cite{fall93}).
	
As conjectured earlier, the high observed spin temperatures of DLAs thus seem 
to be a consequence of their having a higher fraction of the WNM than is found 
for the Galaxy. The fact that the two lowest redshift known DLAs have high $T_s$
means that this cannot be due to evolutionary effects. The possibility that it
is due to a selection effect (i.e. a bias against highly obscured lines of sight) 
cannot, however, be ruled out. Such a bias can only be properly quantified by 
searches for DLAs against bright radio sources, regardless of their optical 
magnitude. Several such searches are currently under way (see e.g. Ellison 2000)
and an observational determination of the importance of dust obscuration should
be available in the near future.

\begin{acknowledgement}
We are grateful to Phil Perillat and Chris Salter for their help with 
observations and data reduction and to Chris for detailed comments 
and suggestions on an earlier draft of this paper. We also thank Frank Briggs 
for illuminating discussions on the ``double-switched'' mode of observation.
The NAIC is operated by Cornell University under a cooperative agreement 
with the NSF. The GMRT observations presented in this paper would
not have been possible without the many years of dedicated effort put in
by the GMRT staff to build the telescope.

\end{acknowledgement}

\end{document}